\RequirePackage{lineno}
\RequirePackage{lineno}
 \documentclass[prd,twocolumn,preprintnumbers,superscriptaddress,nofootinbib]{revtex4}
\usepackage [english]{babel}
\usepackage [autostyle, english = american]{csquotes}
\MakeOuterQuote{"}
\usepackage{tabularx} 
\usepackage[title]{appendix}
\usepackage{graphicx}  
\usepackage{dcolumn}   
\usepackage{bm}        
\usepackage{amssymb}   
\usepackage{multirow}
\usepackage{lineno}
\usepackage[hyperfootnotes=false]{hyperref}
\begin{document}
\newcommand{\beq}{\begin{equation}}
\newcommand{\eeq}{\end{equation}}
\newcommand{\de}{\delta}
\newcommand{\di}{\displaystyle}
\newcommand{\ep}{\epsilon}
\newcommand{\ga}{\gamma}
\newcommand{\Ga}{\Gamma}
\newcommand{\la}{\lambda}
\newcommand{\om}{\omega}
\newcommand{\si}{\sigma}
\newcommand{\ve}{\varepsilon}
\newcommand{\vp}{\varphi}
\newcommand{\dmbarbest} {{(3.36^{+0.46}_{-0.40} {(stat.)}\pm0.06{(syst.)})\times 10^{-3} eV^{2}}}
\newcommand{\dmbest}  {(2.32^{+0.12}_{-0.08})\times10^{-3} eV^{2} }
\newcommand{\pk}{{k \cdot p}}
\newcommand{\pkprime}{{k' \cdot p}}
\newcommand{\kq}{{k \cdot q}}
\newcommand{\pq}{{q\cdot p}}
\newcommand{\W}{{p'}}
\newcommand{\qW}{{q \cdot p'}}
\newcommand{\Wq}{{q \cdot p'}}
\newcommand{\pW}{{p\cdot p'}}
\newcommand{\GeV}{\; {\mathrm{GeV}}}
\newcommand{\cm}{\; {\mathrm{cm}}}
\newcommand{\D}{\displaystyle}
\newcommand{\nuwro}{\textsc{NuWro}}
\newcommand{\pythia}    {{\sc{pythia}}}
\newcommand{\achilles}  {{\sc{achilles}}}
\newcommand{\minerva}{MINER$\nu$A}
\newcommand{\ttbs}{\char'134}
\newcommand{\integ}{\iint}
\newcommand{\FA}{${\cal F}_A$}
\newcommand{\fa}{${\cal F}_A(q^2)$}
\newcommand{\gepQ}{$G_{Ep}(q^2)$}
\newcommand{\nubar}[0]{$\overline{\nu}$}
\newcommand{\gep}{G_{Ep}}
\newcommand{\gmp}{G_{Mp}}
\newcommand{\gmn}{G_{Mn}}
\newcommand{\gmpmu}{G_{Mp}/\mu_{p}}
\newcommand{\gepK}{G^{Kelly}_{Ep}}
\newcommand{\gmpK}{G^{Kelly-upd}_{Mp}}
\newcommand{\gen}{G_{En}}
\newcommand{\gmnmu}{G_{Mn}/\mu_{n}}
\newcommand{\gepnew}{G_{Ep}^{new}}
\newcommand{\gmpnewmu}{G_{Mp}^{new}/\mu_{p}}
\newcommand{\gmpKmu}{G^{Kelly-upd}_{Mp}/\mu_{p}}
\newcommand{\gennew}{G_{En}^{new}}
\newcommand{\gmnnewmu}{G_{Mn}^{new}/\mu_{n}}
\newcommand{\numu}{\nu_{\mu}}
\newcommand{\muminus}{\mu^{-}}
\newcommand{\muplus}{\mu^{+}}
\newcommand{\numubar}{\overline{\nu}_{\mu}}
%
%
\newcommand{\carbon}{\rm ^{12}C}
\newcommand{\oxygen}{\rm ^{16}O}
\newcommand{\deuteron}{\rm ^{2}H}
\newcommand{\hydrogen}{\rm ^{1}H}
\newcommand{\Hefour}{\rm ^{4}He}
\newcommand{\lead}{\rm^{208}Pb}
\newcommand{\Hethree}{\rm ^{3}He}
\newcommand{\neon}{\rm ^{20}Ne}
\newcommand{\aluminum}{\rm^ {27}Al}
\newcommand{\argon}{\rm ^{40}Ar}
\newcommand{\iron}{\rm ^{56}Fe}
\newcommand{\genie}{$\textsc{genie}$}
\newcommand{\qv}{$\bf |\vec q|$}
\newcommand{\rlqe}{${\cal R}_L^{QE}(\bf q,\nu)$ }
\newcommand{\rtqe}{${\cal R}_T^{QE}(\bf q,\nu)$ }
\newcommand{\rltot}{${\cal R}_L(\bf q, \nu)$ }
\newcommand{\rttot}{${\cal R}_T(\bf q, \nu)$ }

\newcommand{\Rochester}{Department of Physics and Astronomy, University of Rochester, Rochester, NY  14627, USA}
\newcommand{\JLAB}{Thomas Jefferson National Accelerator Facility, Newport News, VA 23606, USA}
\newcommand{\Poland}{Institute of Theoretical Physics, University of Wroc\l aw, plac Maxa Borna 9, 50-204, Wroc\l aw, Poland}
\newcommand{\Israel}{School of Physics and Astronomy, Tel Aviv University, Israel}
\hspace{6.1in} \mbox{{Nufact24-2}}
%
%

\title{Summary of Global Extraction of the  $\rm^{12}C$ Nuclear Electromagnetic Response Functions and Comparisons to Nuclear Theory and Neutrino/Electron Monte Carlo Generators\\
"Contribution to the 25th International Workshop on Neutrinos from Accelerators"}
%
%

\author{Arie~Bodek}
\affiliation{\Rochester}
\email{bodek@pas.rochester.edu}
   \author{M.~E.~Christy}
\affiliation{\JLAB}
\email{christy@jlab.org}
\author{Zihao Lin}
\affiliation{\Rochester}
\email{zlin22@ur.rochester.edu}
\author{Giulia-Maria  Bulugean}
\affiliation{\Rochester}
\email{gbulugea@ur.rochester.edu}
\author{Amii Matamoros Delgado }
\affiliation{\Rochester}
\email{amatamor@u.rochester.edu}
\date{\today}


\begin{abstract}
\begin{center} (presented by Zihao Lin)\end{center}
We present a  brief  report (at the Nufact-2024 conference) summarizing a  global extraction of the ${\rm ^{12}C}$ longitudinal  (${\cal R}_L$) and transverse (${\cal R}_T$) nuclear electromagnetic response functions from  an analysis of all available electron scattering data on carbon.  Since the  extracted response functions cover a large kinematic range  they can be readily used for comparison to theoretical predictions as well as  validation  and tuning Monte Carlo  (MC) generators for electron and neutrino scattering experiments.  Comparisons to several theoretical approaches and MC generators are given in detail in  arXiv:2409.10637v1 [hep-ex].  We find that among all the theoretical models that were investigated,  the  ``Energy Dependent-Relativistic Mean Field'' (ED-RMF) approach  provides the best description of both the Quasielastic (QE) and {\it nuclear  excitation} response functions  (leading to single nucleon final states) over all values of  four-momentum transfer.    The QE  data are also well described by the   "Short Time Approximation Quantum Monte Carlo" (STA-QMC) calculation which  includes both single and two nucleon final states  which presently is only  valid for momentum transfer $0.3<{\bf q} < 0.65$  GeV and does not include  nuclear excitations.  
An analytic extrapolation of STA-QMC to lower $\bf q$ has been implemented in the GENIE MC generator for $\rm^{4}He$  and  a similar extrapolation for ${\rm ^{12}C}$ is under development.  STA validity for  ${\bf q} >$ 0.65 GeV requires the implementation of  relativistic corrections.    Both approaches have the added benefit that the calculations are also  directly applicable to the same kinematic regions for neutrino scattering.   In addition we also report on a universal fit to all electron scattering data that can be used in lieu of experimental data for validation of Monte Carlo generators (and is in the process of being implemented in GENIE).
\end{abstract}
 \vspace{-5pt}

\maketitle
%
%
Electron scattering cross sections on nuclear targets are completely  described by longitudinal (${\cal R}_L$)  and transverse (${\cal R}_T$) nuclear electromagnetic response functions. Here  ${\cal R}_L$  and  ${\cal R}_T$  are functions of the energy transfer $\nu$ (or excitation energy $E_x$) and the  square of the 4-momentum transfer $Q^2$ (or  alternatively the  3-momentum transfer $\bf q$). Theoretical calculations
%
of ${\cal R}_L({\bf q}, \nu)$  and  ${\cal R}_T({\bf q}, \nu)$  can be tested by comparing the predictions to experimental data.   We have performed an extraction of electron scattering response functions as functions of (${\bf q}, \nu$), as well as  ($Q^2, \nu)$ from an analysis of all available inclusive electron scattering  cross section data for $\rm^{12}C$.

With  the advent of  DUNE  (Deep Underground Neutrino Experiment), 
 next generation neutrino oscillation experiments aim to search for  CP violations in neutrino oscillations.  Therefore, current neutrino MC generators need to be validated and tuned over the complete range of relevance in $Q^2$ and $\nu$ to provide a better description of  the cross sections for electron and neutrino interactions.  Given the nuclear physics common to both electron and neutrino scattering from nuclei, extracted response functions from electron scattering spanning a large range of  $Q^2$ and $\nu$  also provide a powerful tool for validation and tuning of neutrino  Monte Carlo (MC) generators (run in electron scattering mode). 

Comparisons  of the extracted  ${\cal R}_L$ and ${\cal R}_T$ to several theoretical approaches and  MC generators  are given in detail in \cite{Bodek:2024mrp}.
 These include:  ``Green's Function Monte Carlo'' (GFMC),
`Energy Dependent-Relativistic Mean Field'' (ED-RMF)~\cite{Franco-short} (recently implemented in the  {{\sc{neut}}} MC generator) 
, "Short Time Approximation Quantum Monte Carlo" (STA-QMC)~\cite{Andreoli-short} and "Correlated Fermi Gas". 
Also the     \nuwro{}, 
 \achilles~ (A CHIcago Land Lepton Event  Simulator), 
 and {{\sc{genie}}}~\cite{Andreopoulos:2009rq}  MC generators.  In this summary we focus on comparisons to   ED-RMF,  STA-QMC and GFMC.
%
\begin{figure*}[ht]
\begin{center}
\includegraphics[width=2.3in,height=1.6in]{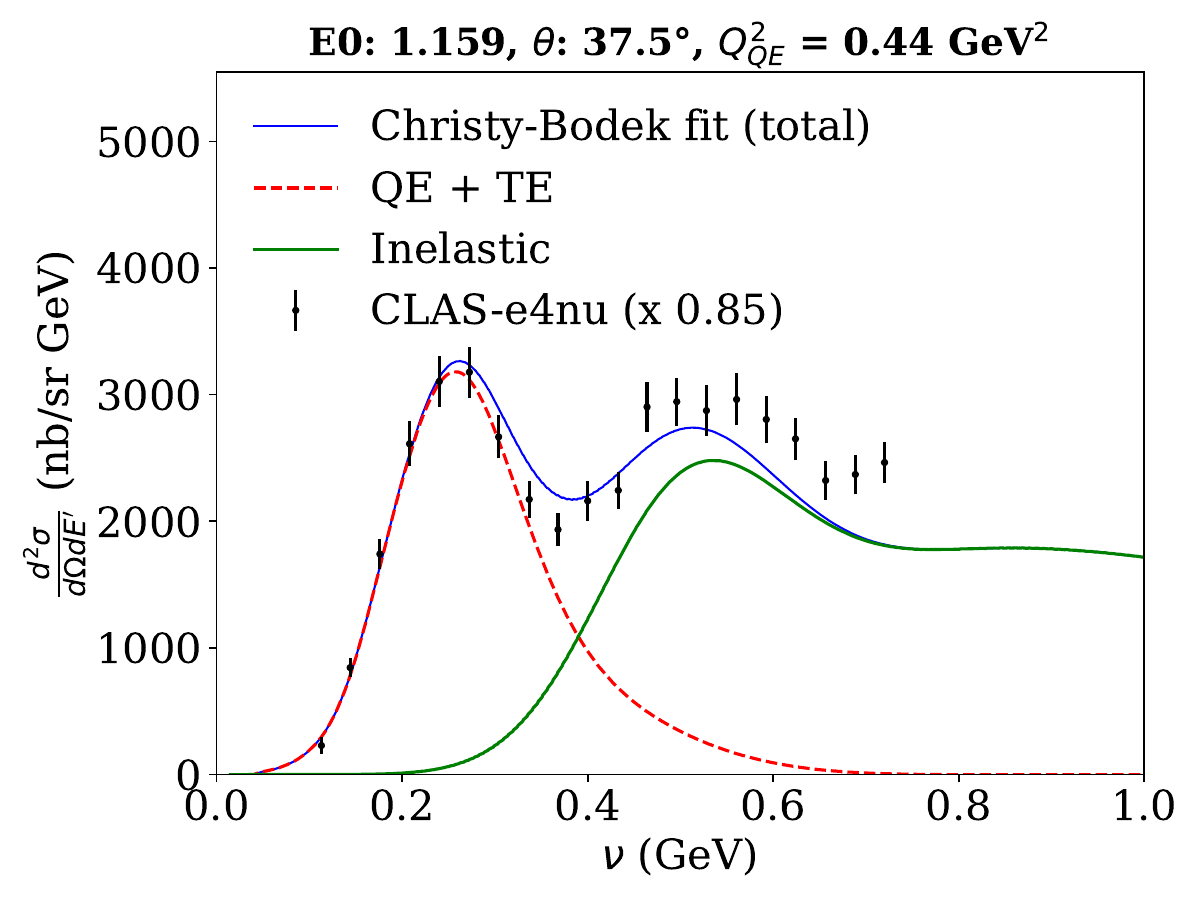}
\includegraphics[width=2.3in,height=1.6in]{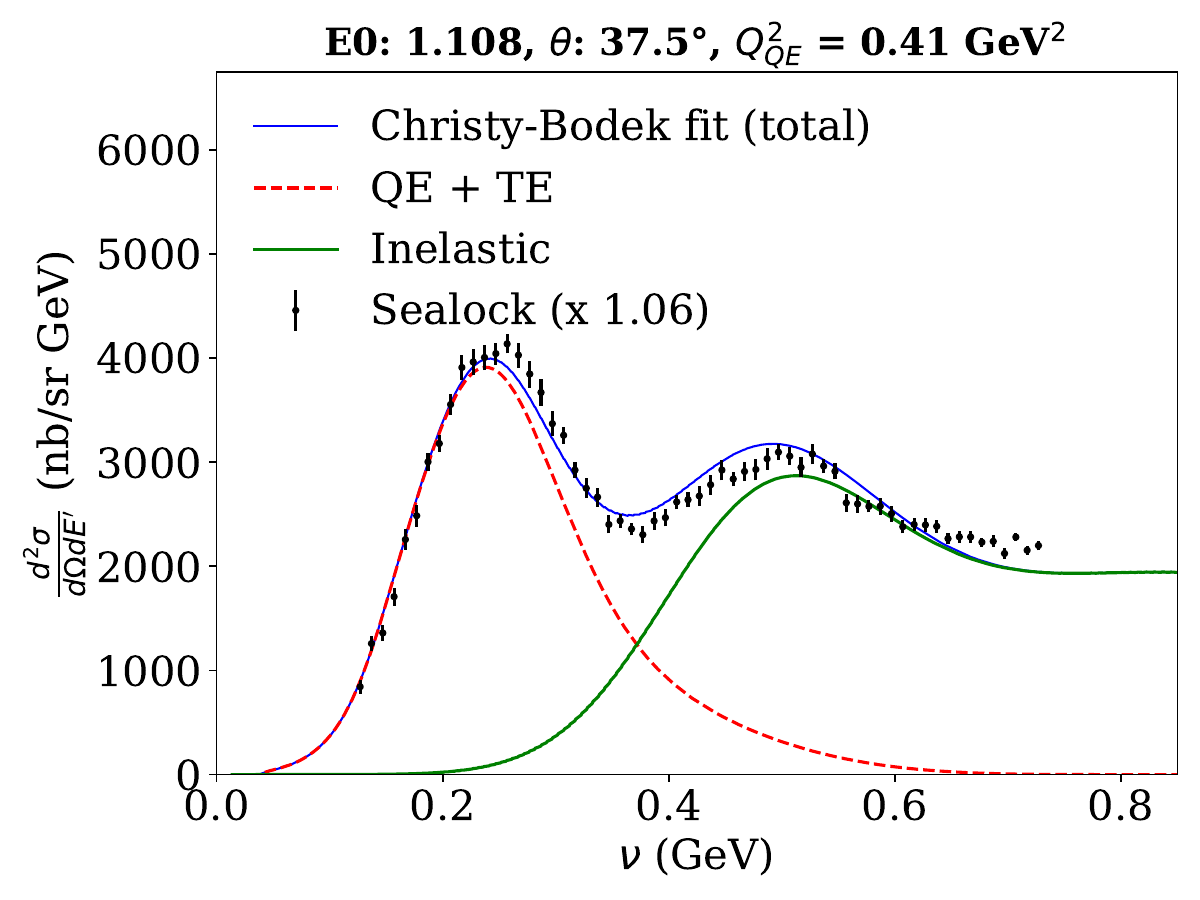}
\includegraphics[width=2.3in,height=1.6in]{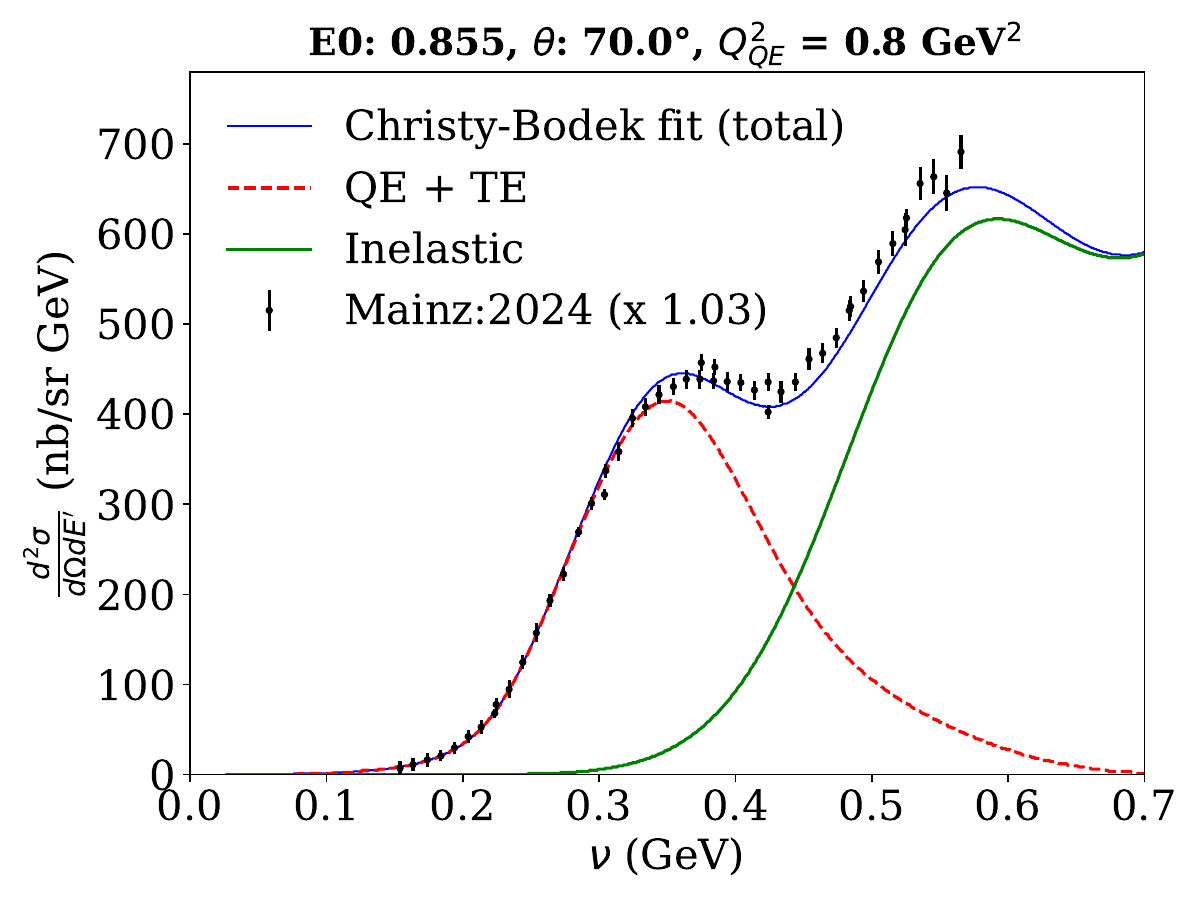}
\includegraphics[width=2.3in,height=1.6in]{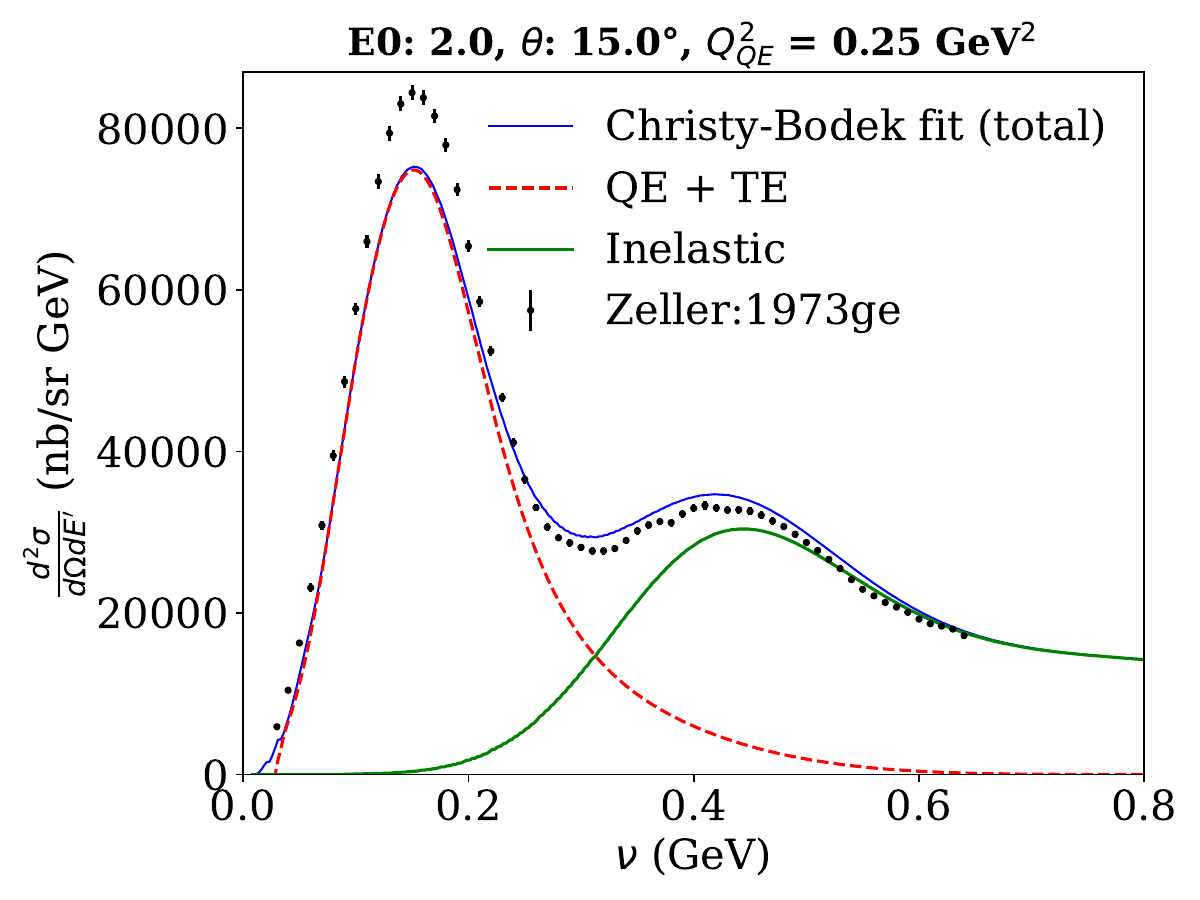}
\includegraphics[width=2.3in,height=1.6in]{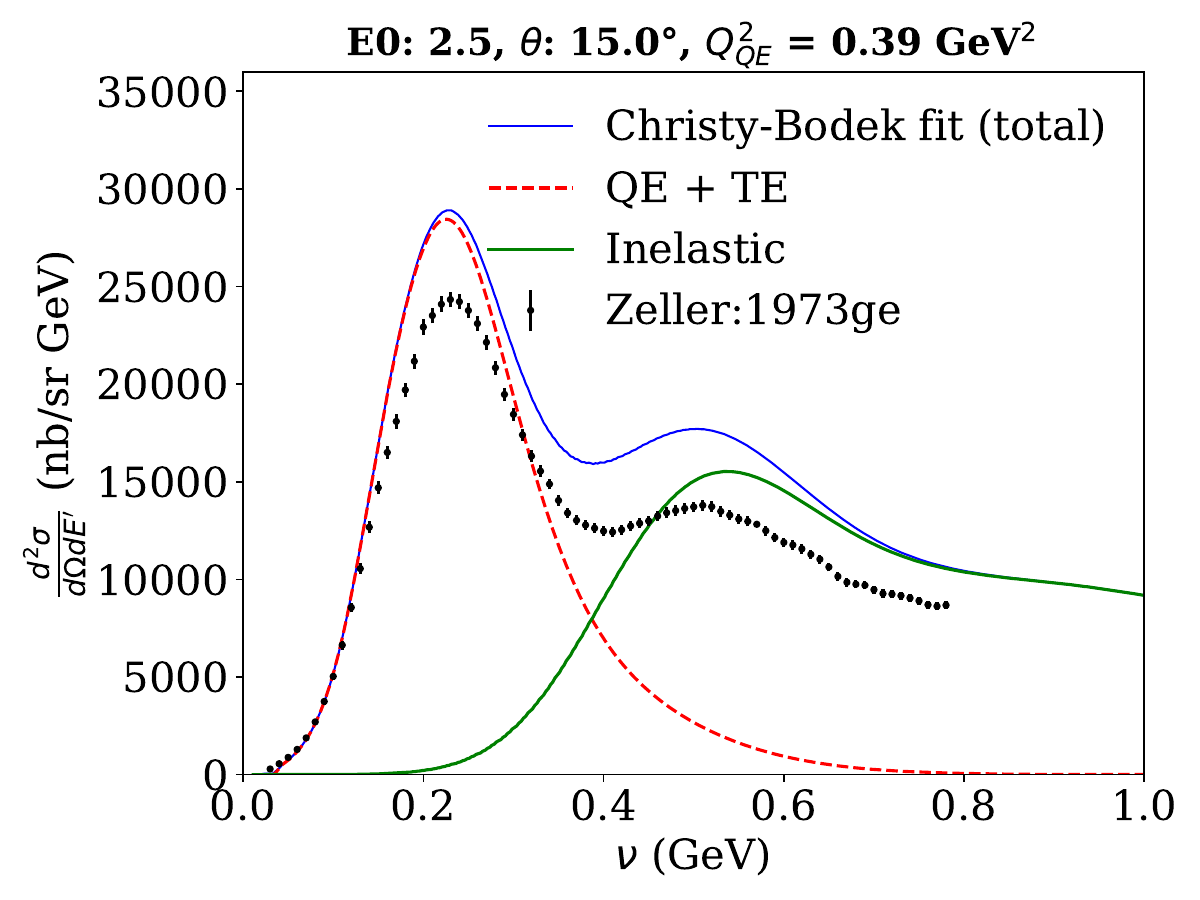}
\includegraphics[width=2.3in,height=1.6in]{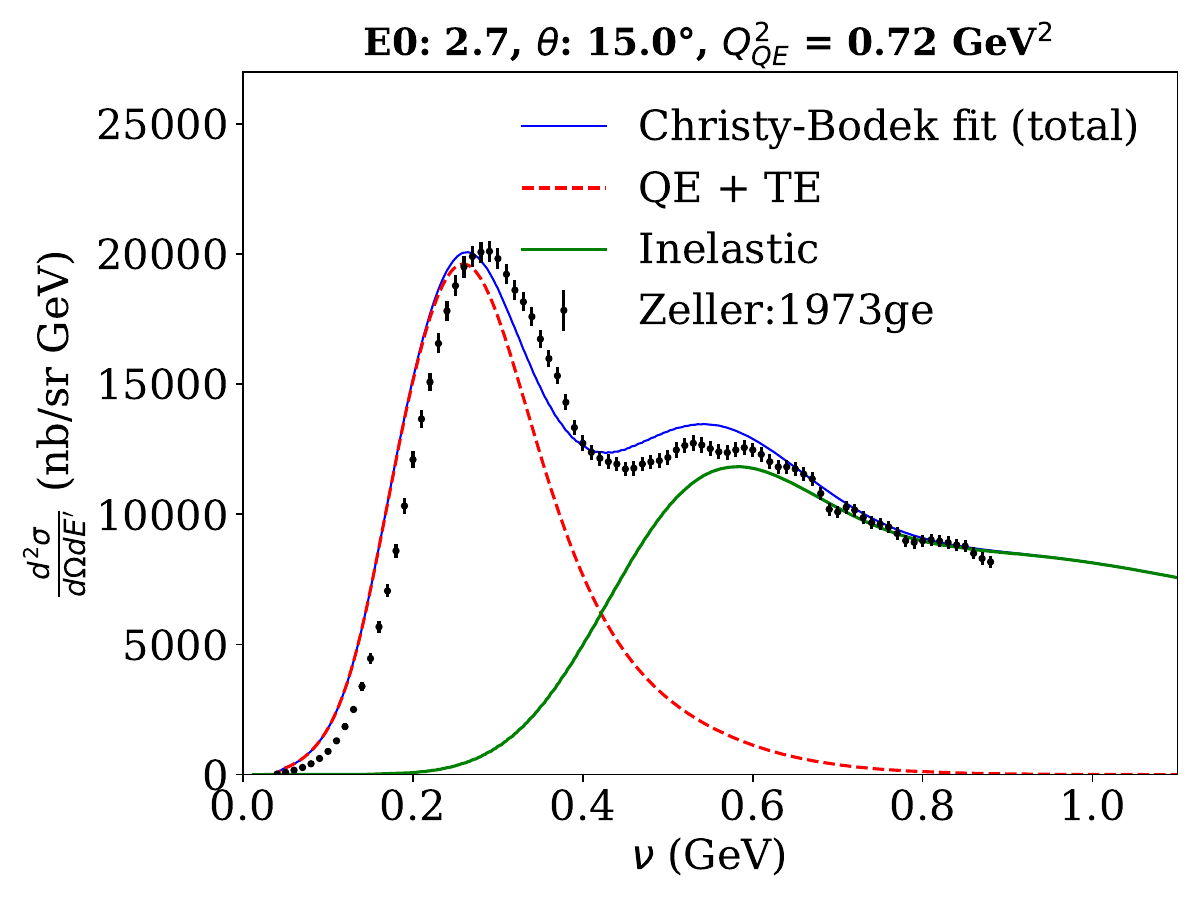}
 \vspace{-12pt}
\caption{Comparison of the Christy-Bodek  fit to measurements. (a)  A new CLAS-e4nu:2023 cross section\cite{Isaacson:2023} measurement (which was not included in the fit) at E$_0$=1.159 GeV and $\theta$=37.45$^0$ (multiplied by 0.85). (b) A previous Sealock:1989 measurement \cite{Sealock:1989nx}  at 1.108 GeV and $\theta$=37.5$^0$ (multiplied by 1.06)  (c) A  new Mainz:2024\cite{Mihovilovic:2024ymj} measurement at 0.855 GeV and $\theta$=70$^0$ (multiplied by 1.03)  (d-f) Zeller:1973 measurements\cite{Zeller:short}  for 2.0, 2.5 and 2.7 GeV at $\theta$=37.5$^0$ (are inconsistent with world data).}
\label{Fig1}
\end{center}
 \vspace{-28pt}
\end{figure*}
%
We briefly describe the steps in the extractions of ${\cal R}_L({\bf q}, \nu)$  and  ${\cal R}_T({\bf q}, \nu)$  from all electron scattering data on $^{12}C$:
We use the Christy-Bodek universal fit~\cite{Bodek:2022gli,Bodek:2023dsr}  to all inclusive electron scattering data on $\rm^{12}C$ (and other nuclear targets) to extract the relative normalizations between different experiments and remove  a few data sets that are inconsistent with all  the other measurements. 
The relative normalizations of all experiments that were used in the fit  are within a few percent of each other\cite{Bodek:2024mrp}.  The fit can then  be used in lieu of experimental data for the testing and validation of new nuclear models and also extract the relative normalization and test the consistency of future electron scattering cross section measurement with previous data. We are  working with  the  {{\sc{genie}}}  group (Joshua Barrow) on the implementation of  the fit as an option in  {{\sc{genie}}}. 

For  example, a comparison of the fit to new CLAS-e4nu:2023 cross section\cite{Isaacson:2023} measurement (which was not included in the fit) at E$_0$=1.159 GeV and $\theta$=37.5$^0$ is shown  in Fig. \ref{Fig1}(a). The normalization of this data relative to previous data is 0.85.   A previous Sealock:1989 measurement \cite{Sealock:1989nx}  at 1.108 GeV and $\theta$=37.5$^0$ is shown in Fig. \ref{Fig1}(b)  (normalization of 1.06), and  a new Mainz:2024\cite{Mihovilovic:2024ymj}  measurement at
 0.855 GeV and $\theta$=70$^0$  is shown in Fig. \ref{Fig1}(c) (normalization of 1.03). Figures \ref{Fig1} (d-f) show the Zeller:1973 measurements\cite{Zeller:short} 
  for 2.0, 2.5 and 2.7 GeV at $\theta$=37.5$^0$ (no normalization).  These measurements were excluded from the universal fit because of inconsistent normalizations between the  three energies,  and the shift in $\nu$ in the 2.7 GeV Zeller data.
 
\begin{figure*}
\includegraphics[width=7.0 in, height=7.in] {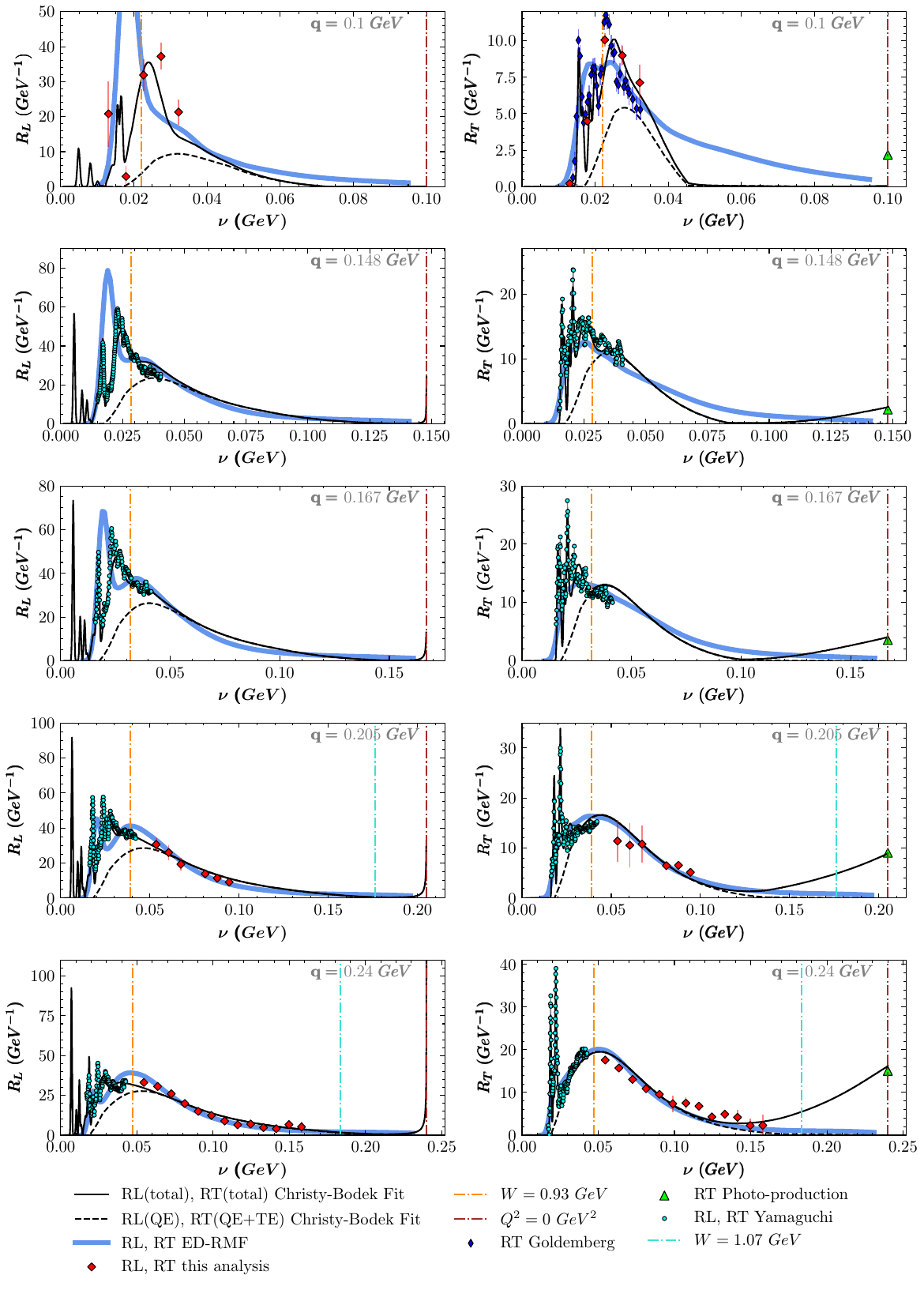}
\caption{Our extractions of ${\cal R}_L$ and ${\cal R}_T$ for  $\bf q$ values of 0.10, 0.148, 0.167, 0.205, and  0.240 GeV versus $\nu$.  In the nuclear excitation region we  also show  measurements from Yamaguchi:1971~\cite{Yamaguchi:1971ua}, and  ${\cal R}_T({\bf q}=0.01)$ GeV extracted  from cross sections at 180$^{\circ}$ (Goldemberg:64 and  Deforest:65). 
The values of   ${\cal R}_T({\bf q}=\nu)$ are extracted from photo-absorption data. In all the plots the  universal fit for the total  (from all processes)  \rltot and \rttot is the solid black line and  the QE component  (including  Transverse Enhancement) of the universal fit  is the dotted line. 
 The thick blue lines are the predictions of the ED-RMF calculations  (which include nuclear excitations).
 }
\label{Fig2}
\end{figure*}

\begin{figure*}
\includegraphics[width=7.0 in, height=8.3in] {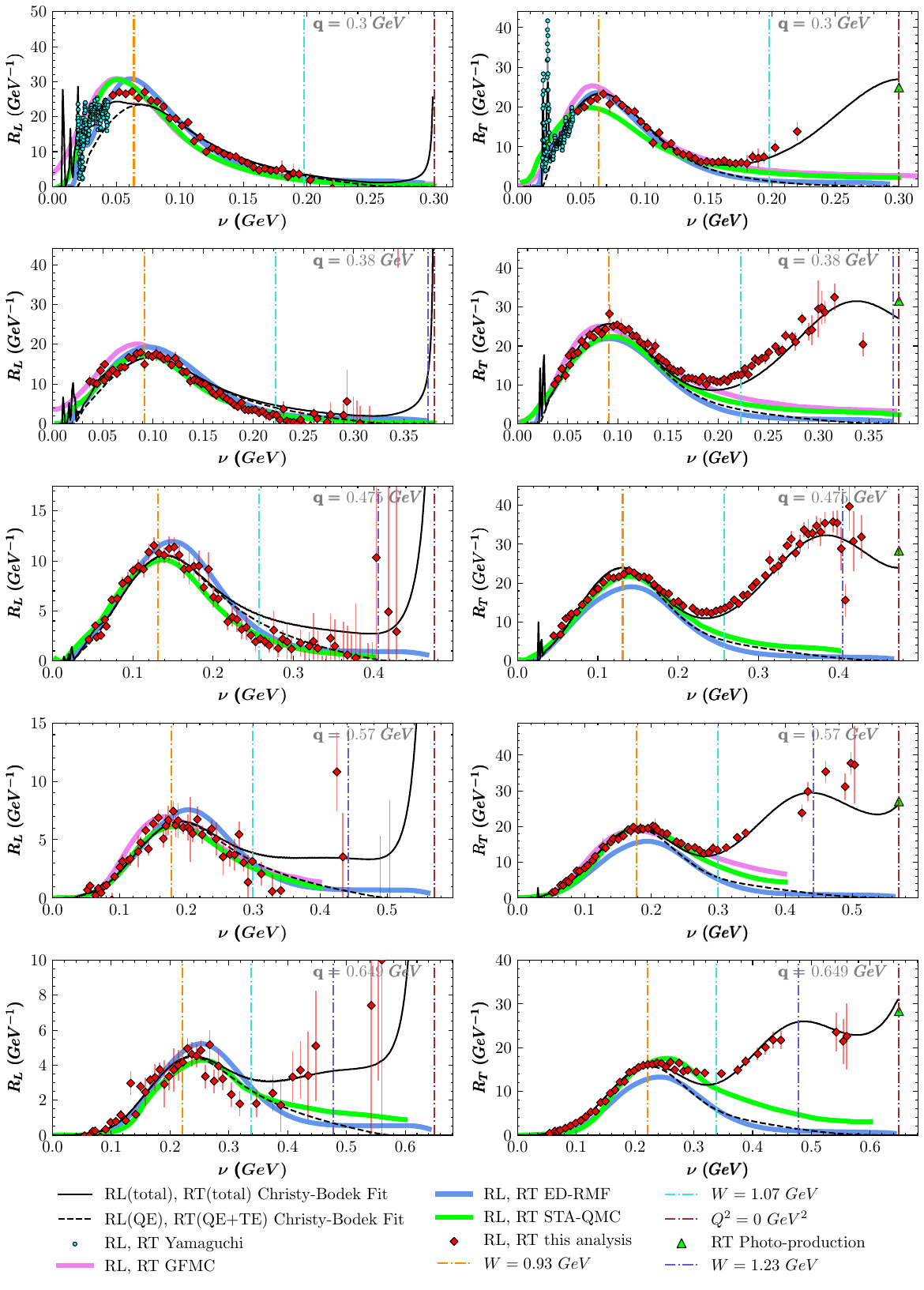}
\caption{Same as Fig. \ref{Fig2} for ${\bf q}$ values of   0.30, 0.38, 0.475, 0.57, and 0.649 GeV versus $\nu$.  the thick blue lines are the predictions of ED-RMF,the thick light green  lines are the predictions of STA-QMC and the  thick pink lines are the prediction of GFMC. }
\label{Fig3}
\end{figure*}

 The extractions of ${\cal R}_L$  and ${\cal R}_T$ shown in Figs \ref{Fig2} and \ref{Fig3}  are done via a Rosenbluth separation of all available  data in bins of $\bf q$ and $\nu$. The universal fit is essential since it is used for bin centering corrections. The values of   ${\cal R}_T({\bf q}=\nu)$ are extracted from photo-absorption data. The universal fit for the total  (from all processes)  \rltot and \rttot is the solid black line and  the QE component  (including  Transverse Enhancement) is the dotted line. 
ED-RMF (thick blue line)  provides the best description of both the Quasielastic (QE) and {\it nuclear  excitation} response functions  (leading to single nucleon final states only) over all values of  four-momentum transfer.   
 The QE  data are also well described by STA-QMC (thick green line) which  includes both single and two nucleon final states  but is only  valid  for   $0.3<{\bf q} < 0.65$  GeV and does not include  nuclear excitations.   An analytic extrapolation of STA-QMC to lower $\bf q$ has been implemented in the GENIE MC generator for $\rm^{4}He$  and  a similar extrapolation for ${\rm ^{12}C}$ is under development. STA validity for  ${\bf q} >$ 0.65 GeV requires the implementation of  relativistic corrections.  Both approaches have the added benefit that the calculations are also  directly applicable to the same kinematic regions for neutrino scattering.   GFMC (pink line) is very CPU intensive and cannot provide predictions for $\bf q < $ 0.3 GeV.  
 
  Research supported  in part by   the Office of Science, Office of Nuclear Physics under contract DE-AC05-06OR23177 (Jefferson Lab) and by  the U.S. Department of Energy under University  of Rochester grant  DE-SC0008475.

%

%
%
\bibliographystyle{apsrev4-1}
\bibliography{12C_Nufact_24}
\end{document}